\documentclass[copyright,creativecommons,noncommercial]{eptcs}
\providecommand{\event}{ICLP 2026} 
\usepackage{iftex}

\ifpdf
  \usepackage{underscore}         
  \usepackage[T1]{fontenc}        
\else
  \usepackage{breakurl}           
\fi

\usepackage{amsmath}
\usepackage{graphicx}
\usepackage{multirow}
\usepackage{hyperref}
\usepackage{listings}
\usepackage{orcidlink}

\title{ASPIC: Proof-of-Concept ASP to Picat Transpiler}
\author{Cristian Grozea\orcidlink{0000-0001-6393-1919} 
\institute{Fraunhofer FOKUS\\ Berlin, Germany}
\email{cristian.grozea@fokus.fraunhofer.de}
\and
Marius Popescu\orcidlink{0000-0003-4158-3911} 
\institute{University of Bucharest\\
Bucharest, Romania}
\email{popescunmarius@gmail.com}
}

\newcommand{\titlerunning}{ASPIC: ASP to Picat Transpiler}
\newcommand{\authorrunning}{C. Grozea, M. Popescu}

\begin{document}

\maketitle

\begin{abstract}
This article presents ASPIC, a new proof-of-concept library that converts extended syntax ASP--Core--2 programs to Picat predicates  that can be solved right away with the integrated Picat SAT solver, or embedded in larger Picat programs (``ASP in Picat''), and that can in turn make use of various Picat predicates and functions (``Picat in ASP'').
The first tests show compatibility with clingo, on programs lacking positive loops and when the special Picat features are not used. With the embedded Picat, it touches the application field of clingcon as well, by being able to efficiently model with both ASP atoms and with finite-domain variables, but goes beyond that by being able to model also non-linear constraints.
\end{abstract}


\section{Introduction}
Picat~\cite{zhou2013user}, is a modern multi-paradigm programming language. It supports logic programming (largely Prolog-compatible), imperative programming (e.g. if then else, foreach loops), functions (including user-defined ones) and functional programming elements (e.g. map), constraint satisfaction and optimisation capabilities (with integrated CP solver, integrated SAT solver, possibility to call external MIP solvers). It offers convenient tabling, turning many exponential-time recursive predicate definitions into polynomial-time dynamic programming ones with a single annotation.

One notable paradigm that Picat has not supported so far is Answer Set Programming. This article explores ways to directly and automatically translate Answer Set Programming (ASP) code to Picat. The outcome of this exploration is a Picat library named ASPIC, available online at \url{https://github.com/picat-lang/lib_ext/tree/main/aspic}. 

ASP is a declarative paradigm characterised by a very compact standardised syntax described in \cite{calimeri2020asp} and sound definition of semantics, based on non-monotonic logic -- see \cite{lifschitz2010thirteen}. The solutions to the ASP programs (when they exist) are named stable models, or answer sets.

Although the ASP programs can contain variables, usually a first phase in finding a stable model is to ``ground'' the program, that is to substitute every variable in every rule and constraint with every possible value and consider the now variable-free rules resulting. This grounding can lead to a high memory usage and can take a fairly long time, issues that together bear the name ``the grounding bottleneck''.

Several efficient ASP solvers exist, such as clingo~\cite{gebser2011potassco}, which combines a grounder named gringo with a solver for grounded problems named clasp. Clingo is very efficient, being for example able to solve within minutes the N-Queens problem for 5000 queens. A new solver for ASP is thus not necessarily needed. What the ASP solvers seem to lack (although present in some ASP environments such as ASPchef~\cite{alviano2023introducing}) is a way to cover other tasks such as data preprocessing or postprocessing steps without resorting to completely different programming languages, such as Python and Lua (in the case of the clingo integrations).

Therefore, in contrast with those purely procedural languages we think that a tight integration between ASP and a language with a solid logic programming foundation, such as Picat, raises interesting possibilities, beyond simply introducing a new SAT-based ASP solver.

An additional aspect where ASPIC improves over the state of the art is by being able to include nonlinear CP constraints into the ASP model, which is more than clingcon and clingo-dl described in \cite{janhunen2017clingo}, which support (pseudo-)linear constraints over finite-domain integers, and difference logic over integers/reals, respectively.
\paragraph {Related work:}
We focus here on other transpilers from ASP to logic or constraint solving languages, and on libraries adding ASP capabilities to those targets, and not on stand-alone ASP solvers, some of which, like the two systems mentioned above, clingcon and clingo-dl also have the possibility to embed finite-domain constraints into ASP programs.

The system s(CASP) was written for Ciao Prolog and ported later to SWI-Prolog~\cite{wielemaker2021s}, although with its goal-oriented approach, s(CASP) is closer to Prolog than the regular ASP. There are two systems most closely related to our work: ASP2AR~\cite{10.5555/2010192.2010243} where ASP is translated into what basically amounts to a solver based on Action Rules (AR) that implement the event-based propagation of the constraints; and ASP-PROLOG~\cite{elkhatib2005integrating} where an ASP solver is integrated inside Prolog, allowing thus both preprocessing of the input and postprocessing (e.g. for filtering) using Prolog code.  
Another system that converts ASP programs to constraint programming models is ASP-FZN~\cite{eiter2025asp}, where the ASP programs are converted to FlatZinc, a strict subset of the constraint modelling language MiniZinc, opening thus the possibility to leverage several SAT and MIP solvers to solve ASP problems.
Concerning the richness of the constraint programming possible in ASPIC due to the integration with Picat, only the inductive predicates for MiniZinc described in~\cite{aziz2013inductive} would be as rich, with the caveat that this experimental extension of MiniZinc seems to be unavailable in the current MiniZinc. We could also not detect any replacement or evolution of it in the MiniZinc documentation.

\section{Approach}
The approach proposed here is to use natural mappings between the ASP objects and native Picat constructs.
The transformation will be illustrated on the minimal ASP program 
from \url{https://personal.utdallas.edu/~gupta/nfm-ex/scasp-manual.pdf} given in Listing~\ref{lst:path}.
\begin{lstlisting}[float=t,language=prolog,caption=Minimal ASP program,label=lst:path]
edge(a,b).
edge(b,c).
edge(b,d).
edge(c,e).
edge(d,a).
path(X,Y) :- edge(X,Y).
path(X,Y) :- edge(X,Z),path(Z,Y).  
\end{lstlisting}

The ASP atoms (variable-free after grounding) correspond naturally to binary variables, as they can either belong or not to a stable model. Picat has efficient associative data structures named maps (similar to the dictionaries in Python and to hash maps in C++), that can accept arbitrary grounded ASP atoms as keys and can be used to store the binary constraint programming (CP) variables associated to the atoms.

The path definition program in Listing~\ref{lst:path} has $U=\{a,b,c,d,e\}$ as the set of values occurring in atoms as arguments for the functors \verb|edge| and \verb|path|. This means that the total number of atoms implied by the rules with variables is 50: 25 atoms of the form \verb|path(X,Y)| with arbitrary \verb|X| and \verb|Y| in $U$, and further 25 atoms of the form \verb|edge(X,Y)|. 
\begin{lstlisting}[caption=Simple ASP Rule, label=lst:ep,float=t]
path(X,Y) :- edge(X,Y).    
\end{lstlisting}
When grounding the rule in Listing~\ref{lst:ep}    
ASPIC produces 25 rules of similar form, one for every combination of X and Y, as shown in Listing~\ref{lst:grounded}.
\begin{lstlisting}[float=t,caption={Simple ASP Rule, Grounded},label=lst:grounded]
path(a,a) :- edge(a,a).
path(a,b) :- edge(a,b).
path(a,c) :- edge(a,c).
...
path(e,e) :- edge(e,e).
\end{lstlisting}
Please note that this is a very straight-forward approach used in ASPIC, and that a more advanced grounder like gringo (integrated in clingo) does that in a more efficient fashion, by only using the 5 combinations that actually appear for the functor \verb|edge/2|.

The ASP rules, of the form \verb|head :- body.| correspond after grounding to implication constraints in the CP sense, by stating that whenever the body holds the head of the rule must hold as well, that is $body\rightarrow head.$

The ASP constraints, of the form \verb|:- body.| 
with the meaning that the conjunction of predicates in the body should not hold under the assignments, 
can as well be seen as simple constraints in the CP sense, namely \verb|not(body)|.

With such direct mapping of the ASP rules and constraints to Picat constraints, the code produced by the transpiler is not only human readable but it corresponds in a very obvious way to the original ASP code. This means that this transformation can, for example, be used  as a starting point, by further editing the result in postprocessing if so desired.

Whereas the transformation as described so far produces assignments that are solutions in the Boolean sense, they are not necessarily stable models, as they can lack the requirement of supportedness, that is any atom can be in the stable model only if it's supported by a rule in the ASP program, that is it becomes true only by being implied by elements already in the stable model through a rule.

Towards supportedness, ASPIC uses the Clark completion~\cite{clark1977negation}, which requires turning around all implications and adding the result as new rules. Intuitively, if $head\leftarrow body1$ and $head\leftarrow body2$ are the only two rules that can produce $head$ then the completion adds a rule of the form $head\rightarrow(body1;body2)$. This is being generalised to more than 2 rules for the same head in the expected way.

Note that Clark completion does not guarantee obtaining stable models, as supporting loops can exist for atoms that cannot be derived starting with facts and applying repeatedly rules only on atoms previously proven. The simplest example of such a loop is this:
\begin{lstlisting}[language=prolog]
    p:-p.
\end{lstlisting}
The only stable model of this minimal program is the empty set, but $p \leftrightarrow p$ (which is the outcome of Clark completion here) admits the solution $\{p\}$ as well, which is not a stable model. Stable models are stricter, they require truth to be founded on facts, not circular reasoning.

Although everything is easier to explain and implement on grounded rules and constraints, grounding may be a bottleneck, thus some ASP systems try to use partial grounding or even no grounding at all. ASPIC uses grounding on-the-fly, in order to avoid the program size explosion after conversion -- note that this does not solve the combinatorial explosion of the runtime due to unfiltered exhaustive grounding. Mathematically, the meaning of the rule given in Listing~\ref{lst:ep} is this $$\forall X\in U,\forall Y\in U,edge(X,Y)\rightarrow path(X,Y).$$
where the universal quantification can be expressed in Picat using the {\em foreach} construction: 
\begin{lstlisting}
    foreach(X in U, Y in U) edge(X,Y) #=> path(X,Y) end
\end{lstlisting}
where \verb|#=>| is the Picat CP operator for implication and U is supposed to contain the ``universe'', the set of all values over which the iteration should go (in this case of size 5).
For this translation to function properly in Picat despite being so close to the original ASP, one needs that the Picat construct \verb|edge(X,Y)|, with instantiated (grounded) \verb|X| and \verb|Y| evaluates to the binary CP variable associated with the ASP atom \verb|edge(X,Y)|. Here the availability of user defined functions in Picat is helpful, as one can obtain this effect by defining these functions:
\begin{lstlisting}
path(X1,X2)=aspic_var($path(X1,X2)).
edge(X1,X2)=aspic_var($edge(X1,X2)).   
\end{lstlisting}
where the Picat terms with a \$ in front of the functor are plain terms and no function calls, as per Picat syntax, and \verb|aspic_var| is the function that maps a term to its associated CP variable.

The last rule in Listing~\ref{lst:path} is similarly translated to:
\begin{lstlisting}
foreach(X in U, Y in U,Z in U) 
    edge(X,Z) #/\ path(Z,Y) #=> path(X,Y) 
end    
\end{lstlisting}
where \verb|#/\|~is the constraint conjunction operator and \verb|#=>|~is the constraint implication operator in Picat.

The universe $U$ is derived automatically; for this the ASPIC transpiler makes two passes on the ASP source. The first one gathers and converts all facts and rules, producing the set of terms to be iterated upon, and then the second pass converts the constraints to Picat. Heuristics are used in an attempt to reduce the domain of the variables whenever possible, resulting in tighter \verb|foreach| loops as seen in Listing~\ref{lst:queens5:aspic}.

\subsection{Extending the Syntax and FD Variables}
\begin{lstlisting}[float=t,language=prolog,caption=Non-linear CP Example,label=lst:mini-fd,basicstyle=\small,
morekeywords={foreach,end,in,import,if,then,else,main,cond,all_different,&},
]

is_square(E):-E=0..1024, &cond(S=fd(x,6),S*S#=E,0)&.

\end{lstlisting}
\begin{lstlisting}[float=t,language=prolog,caption=Mixing Picat global constraints and ASP,label=lst:global,numbers=left]
{e(X,Y):X=1..5,Y=1..5,X!=Y}.

&cond((
N=5,
V=[{I,1}:I_in_1..N], %vertices
E=[{I,J,e(I,J)}:I_in_1..N,J_in_1..N], %edges
tree(V,E)),1,0)&. 
\end{lstlisting}

\paragraph{Finite-domain variables:}
In order to facilitate the interaction between the ASP part and the CP part of an ASPIC program, the finite-domain variables are implemented as linear combinations of the binary variables corresponding to ASP predicates with the indicated functor, and with the desired number of bits.
The function $$fd(f,n)=\sum_{i=0}^{n-1}2^i*aspic\_var(f(i))$$ 
makes this link between a functor (here $f$) and the finite-domain (FD) variable of $n$ bits.

\paragraph{Extended syntax:}
To embed Picat code into ASP, we used the ``\&'' delimiter for parts that are to be included verbatim into the generated code.
Those parts should evaluate to Picat expressions that are either constraint values with the domain 0..1 or one of the constants 0 or 1. Thus they can replace an ASP predicate in the generated SAT or CP problem.
This embedding gives access to the full power of the CP modelling in Picat, which allows for polynomial constraints as well, surpassing thus the limitation existent in clingcon to only linear constraints. 

In Listing~\ref{lst:mini-fd} we define a predicate \verb|is_square| that uses a FD variable \verb|S| of 6 bits to constrain E to be the square of \verb|S| by using the Picat CP constraint \verb|E#=S*S|. 
The function \verb|cond(Predicate,YesEx,NoEx)| is a Picat function that implements the so-called computed-if (or conditional expression). It is supported in Picat constraints expressions as well. Here it is used to embed Picat code (the code for defining the variable $S$) into the expression. The predicate can be a conjunction as well, thus more code can be embedded there at once.

\paragraph{Using Picat global constraints in ASP programs:} Picat has a large collection of very efficient global constraints. The Listing \ref{lst:global} shows how to apply those on lists of ASP atoms.
The embedded Picat block (lines 3 to 7), used as a fact, applies the global constraint \verb|tree| to the adjacency matrix defined by the ASP predicate \verb|e|, defined on line 1 of the listing and converted in line 6 to the form Picat expects. Please note the underscores replacing spaces in the embedded Picat, this is now needed as the tokenizer is quite simple and removes all white spaces. This hack will be eliminated in the future.
Thanks to the functions defined by the transpiler for each suitable ASP atom (see line 3 of Listing~\ref{lst:queens5:aspic} for an example), one can refer the ASP atoms inside the embedded Picat code exactly the same way one would do in ASP. 

\section{Examples}

\subsection{N-Queens}
N-Queens is a classical combinatorial puzzle, requiring the placement of $N$ queens on an $N$x$N$ chess-like table such that no two queens are on the same row, column, or diagonal.

The ASP program in Listing~\ref{lst:queens5} is one of the examples offered for trying out clingo online at \url{https://potassco.org/clingo/run/?example}. Please note that it is not the same program on which clingo achieves the amazing performance of solving the problem for 5000 queens within minutes.
\lstset{
         numbers=right,           
         stepnumber=1,                     
         numberfirstline=true
 }

\begin{lstlisting}[float=t,language=prolog,caption=N-Queens,label=lst:queens5,numbers=left]
#const n = 4.
{ q(I,1..n) } == 1 :- I = 1..n.
{ q(1..n,J) } == 1 :- J = 1..n.
:- { q(D-J,J) } >= 2, D = 2..2*n.
:- { q(D+J,J) } >= 2, D = 1-n..n-1.    
\end{lstlisting}

\begin{lstlisting}[float=t,language=prolog,caption=N-Queens transformed through ASPIC,label=lst:queens5:aspic,basicstyle=\small,
morekeywords={foreach,end,in,import,if,then,else,main},numbers=left
]
import sat.
import aspic_runtime_debug.
q(X1,X2)=aspic_var($q(X1,X2)).
main=>main(["sat"]).
main([ASPIC_SEARCH])=>ASPIC_SEARCH:=ASPIC_SEARCH.to_atom(),
aspic_startlearning(),
foreach(I in 1..4)
aspic_holds(aspic_card([q(I,ARANGE1):ARANGE1 in 1..4],1,1)),
end,
foreach(J in 1..4)
aspic_holds(aspic_card([q(ARANGE3,J):ARANGE3 in 1..4],1,1)),
end,
aspic_stoplearning(),
foreach(D in 2..aspic_mul(2,4))
aspic_holds(aspic_not(aspic_conj(
    [aspic_card([q(aspic_diff(D,J),J):J in 1..4],2,[]),1]))),
end,
foreach(D in aspic_diff(1,4)..aspic_diff(4,1))
aspic_holds(aspic_not(aspic_conj(
    [aspic_card([q(aspic_sum(D,J),J):J in 1..4],2,[]),1]))),
end,
aspic_stoplearning(),
H=get_heap_map(aspic_atoms),
foreach(ASPIC_ATOM in H.keys())
if(not aspic_isfdvariable(ASPIC_ATOM))
L=[]
++[1:_ in 1..1,ASPIC_ATOM=$q(I,ARANGE1),I in 1..4,ARANGE1 in 1..4]
++[1:_ in 1..1,ASPIC_ATOM=$q(ARANGE3,J),ARANGE3 in 1..4,J in 1..4]
,L2=aspic_disj(L)
,if(not(L2==1)) then
aspic_holds(aspic_impl(H.get(ASPIC_ATOM),aspic_disj(L))),
end end end,true.
\end{lstlisting}
This is how the automatically generated Picat Listing~\ref{lst:queens5:aspic} corresponds to the original in Listing~\ref{lst:queens5}:
The function definition on line 3 prepares the context such that referring to the binary CP variables associated to the q(X,Y) atoms is done directly with the same syntax as in the ASP original code;
lines 7-9 correspond to the rule on the second line, lines 10-12 to the rule on line 3. The lines 14 to 17 add the constraints from line 4 in the source, whereas the lines 18 to 21 add the one from line 5. The Clark completion occupies the lines 23 to 32, and it's done with the same code for all grounded atoms collected during the direct rules; the reverse implication states that each of those atoms implies the disjunction of the matching rules. Some iterations are eliminated under the assumption that variable matching with the clauses' heads will lead to assignments of the variables used in those.
\subsection{SEND+MORE=MONEY}
This is a classical puzzle (first published by Dudeney in 1924) that requires finding distinct digit values for the 8 letters S,E,N,D,M,O,R,Y, with S and M non zero, such that the addition SEND+MORE=MONEY holds when substituting the letters with the values found and interpreting the sequences of digits as numbers in base 10.
We give two program versions, a pure ASP one in Listing~\ref{lst:money} and one using embedded Picat and FD variables in Listing~\ref{lst:money-fd} - this one is hand-written and has not been generated automatically from the first listing.
\begin{lstlisting}[float=t,language=prolog,caption=Money puzzle,label=lst:money,basicstyle=\small,
morekeywords={foreach,end,in,import,if,then,else,main},
numbers=left
]
n(0..9).
sol(S,E,N,D,M,O,R,Y):-n(S),n(E),n(N),n(D),n(M),n(O),n(R),n(Y),
S!=0,M!=0,
S*1000+E*100+N*10+D+M*1000+O*100+R*10+E=M*10000+O*1000+N*100+E*10+Y,
S!=E,S!=N,S!=D,S!=M,S!=O,S!=R,S!=Y,
E!=N,E!=D,E!=M,E!=O,E!=R,E!=Y,
N!=D,N!=M,N!=O,N!=R,N!=Y,
D!=M,D!=O,D!=R,D!=Y,
M!=O,M!=R,M!=Y,
O!=R,O!=Y,
R!=Y.
\end{lstlisting}

\paragraph{FD variables version:}
\begin{lstlisting}[float=t,language=prolog,caption=Money puzzle -- FD variant,label=lst:money-fd,basicstyle=\small,
morekeywords={foreach,end,in,import,if,then,else,main,cond,all_different},
numbers=left
]
main:-
    &cond((
        S=fd(s,4),
        E=fd(e,4),
        N=fd(n,4),
        D=fd(d,4),
        M=fd(m,4),
        O=fd(o,4),
        R=fd(r,4),
        Y=fd(y,4),
        [E,N,D,O,R,Y]::0..9,
        [S,M]::1..9,
        (1000 * S + 100 * E + 10 * N + D) +
        (1000 * M + 100 * O + 10 * R + E) #=
        (10000 * M + 1000 * O + 100 * N + 10 * E + Y),
        all_different([S,E,N,D,M,O,R,Y])
    ),1,1)&.
\end{lstlisting}
The wrapper on lines 2 and 17 turns Picat code into an expression that can be used as a CP value in the generated program. Please note that the FD variables S,E,N,D,M,O,R,Y are defined on 4 bits each and restricted further to their domains (digits, S and M non-zero). In contrast to clingcon, the predicate \verb|all_different| is not dedicated to this extension towards CP of ASP but is instead one of the many global constraints available in Picat. Other such global constraints model efficiently circuits, acyclic graphs, Hamiltonian cycles, strongly connected components, non-overlapping rectangles, non-overlapping sets of tasks, a graph being a tree, etc.

\section{Benchmarks}
\paragraph{Methodology:} The benchmarking has been performed on the same ASP file when possible. It consisted in repeatedly running the solver under the control of the software ``hyperfine''\footnote{https://github.com/sharkdp/hyperfine} for at least 10 iterations. The mean running time is reported here.

The machine used for benchmarking has two AMD EPYC 9654 96-Core processors and 1.5TB RAM. We chose a powerful machine in order to avoid the performance being limited by the RAM or by the multithreading limitations. Picat uses a single core and therefore also the code converted with ASPIC is executed on a single core, whereas clingo is multithreaded. The software releases used for clingo and picat were the newest available at the time. The operating system was Ubuntu 24.04.

\subsection{N-Queens}
\begin{figure}
    \centering
    \includegraphics[width=0.75\linewidth]{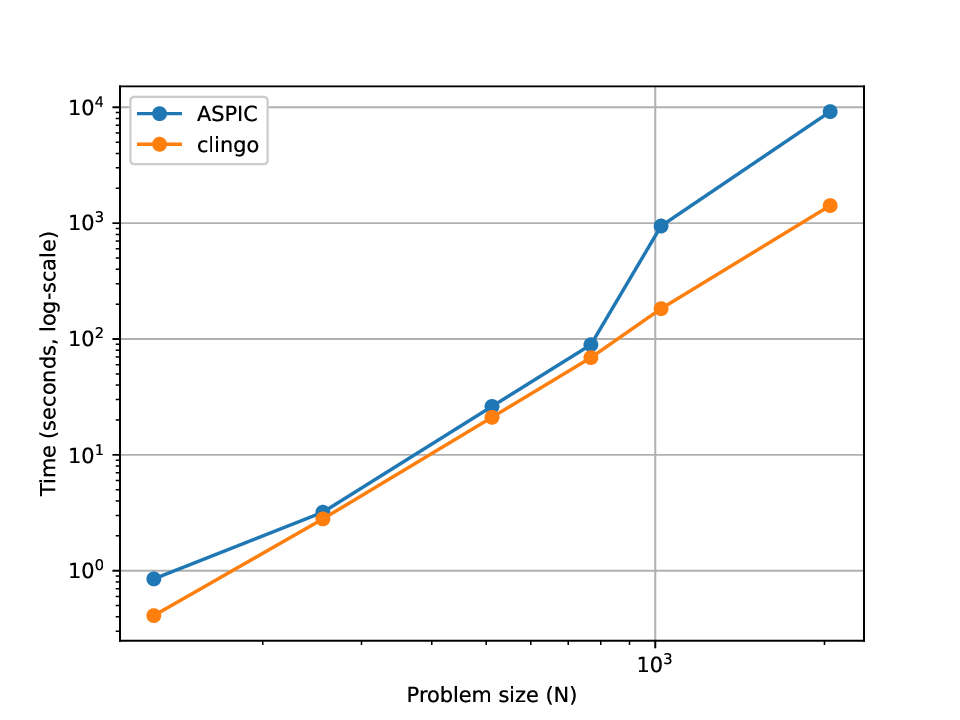}
    \caption{Graphical comparison of the execution times of ASPIC and clingo on the N-Queens problem. Note the increased gap starting with 1024.}
    \label{fig:nqueens}
\end{figure}

The results in Table~\ref{tab:nqueens}, as well as in Figure~\ref{fig:nqueens} are obtained using the ASP program in Listing~\ref{lst:queens5}. The speeds of ASPIC and clingo on this program are quite close for an extended range of problem sizes, but the gap increases when the number of CP variables generated by ASPIC exceeds one million. Clingo has been run single-threaded, as the multithreaded call (using ``-t 64'' that instructs clingo to use 64 threads, the maximum it can use) led to about 10\% longer runtimes.

\begin{table}[ht]
 \centering
 \caption{Comparative Running Times on N-Queens}\label{tab:nqueens}
 {\begin{tabular}{@{\extracolsep{\fill}}lrrrrrr}
   \hline
    Solver& N=128&
     N=256&
     N=512&
     N=768 &
     N=1024 &
     N=2048\\
    \hline
ASPIC (over Picat 3.9\#4)           & 0.85 s & 3.2 s & 26.2 s & 89.3 s & 944.6 s & 9178.9 s\\
clingo 5.8.0    & 0.41 s & 2.8 s & 21.1 s & 69.1 s & 182.8 s & 1418.3 s \\
   \hline
    \end{tabular}}
\end{table}

\subsection{Towers of Hanoi}
\begin{lstlisting}[float=t,caption=Towers of Hanoi in pure ASP,label=lst:hanoi,basicstyle=\small,
morekeywords={foreach,end,in,import,if,then,else,main,cond,all_different},
numbers=left
]
#const n=1024.
#const k=10.

p(1..3). %towers                                                                                                                                                                                                   
x(1..k). %discs                                                                                                                                                                                                    
at(1,1,1..k). %initial                                                                                                                                                                                             
:-not at(n,3,1..k). %final                                                                                                                                                                                         
{at(T,P,X):p(P)}=1:-T=1..n,x(X).
{move(T,P1,P2,X):p(P1),p(P2),at(T,P1,X),P1!=P2}=1:-T=1..n-1.
at(T,P2,X):-T=2..n,p(P2),p(P),move(T-1,P,P2,X).
at(T,P,X):-T=2..n,p(P),at(T-1,P,X),not move(T-1,X).
move(T,X):-x(X),p(P1),p(P2),move(T,P1,P2,X). %projection                                                                                                                                                           
movefrom(T,P,X):-p(P),p(Q),x(X),move(T,P,Q,X). %projection                                                                                                                                                         
moveto(T,Q,X):-p(P),p(Q),x(X),move(T,P,Q,X). %projection                                                                                                                                                           
% cannot place over a smaller one                                                                                                                                                                                  
:-p(P2),moveto(T,P2,X),at(T,P2,Y),Y<X.
% can only move the smallest one                                                                                                                                                                                   
:-p(P),x(X),movefrom(T,P,X),at(T,P,Y),Y<X.

#show move/4.

\end{lstlisting}

The pure ASP implementation of the "Towers of Hanoi" planning puzzle shown in Listing \ref{lst:hanoi} defines the problem as a constraints one: 
there are three rods, and $k$ disks (line 2) of sizes 1..$k$ (line 5) are placed on the first rod (line 6). They are placed in such a way that never a larger disc is placed over a smaller one. In the end, at step $n$, they should all be placed on the rod 3 (line 7). At every time point each of the discs is on a single rod (line 8). At every time step a single disc is moved to another rod (line 9). A disc gets to be at a time step on a rod if either has been moved there at the previous step (line 10) or was there already and has not been moved (line 11).Three projections of the atom move are defined on lines 12 to 14, used to define the constraints: one cannot place a disc in a tower containing a smaller one (line 16) and one can only move the smallest one from a certain rod (line 18).

For this benchmarking we used in addition to the plain clingo invocation also the parameter ``-t 64'' that instructs clingo to use 64 threads, as this led to improvements on this problem. The results are shown in Table \ref{tab:hanoi}. Once the numbers went past the time ASPIC needs to ground this problem (seconds),  it dominated both clingo and multithreaded clingo, being up to almost twice as fast for large instances.

\begin{table}[ht]
 \centering
 \caption{Comparative Running Times on Towers of Hanoi}\label{tab:hanoi}
 {\begin{tabular}{@{\extracolsep{\fill}}lrrrrrr}
   \hline
Solver           &k=7      &k=8      &k=9      &k=10     \\
\hline
clingo          &6.5 s   &44.3 s  &221.6 s &     1746.4 s  \\
clingo -t 64     &\textbf{2.0 s}   &24.7 s  &157.0 s &     1328.9 s  \\
ASPIC           &3.6 s   &\textbf{17.5 s}  &\textbf{119.0 s} &     \textbf{713.2 s}  \\
   \hline
    \end{tabular}}
\end{table}

\subsection{SEND+MORE=MONEY}
The grounding bottleneck for the classical SEND+MORE=MONEY puzzle arises from 
(at least in a naive direct approach like the current one of ASPIC) having to iterate over all possible values 
and get $10^8$ versions 
of the rule starting at line 2 in Listing~\ref{lst:money}.
While clingo's grounder does not literally generate all those $10^8$ versions of the rule, the huge search space still contributes to its runtime: clingo needs 0.756 seconds to solve it on our test machine, 
whereas ASPIC with its explicit unfiltered grounding on-the-fly is hopelessly slow 
thus, within reasonable time, it does not finish grounding it to start solving.

ASPIC solves the FD variant of the problem within 108~ms, compared to clingo's 756~ms for the pure ASP version, which proves once again the advantage of being able to mix ASP programs with CP constraints rather than being a like-for-like comparison.

\section{Current limitations}

At present the proof-of-concept ASPIC implementation uses only Clark completion and therefore, in general, computes models of the completion rather than full stable models. For tight programs (without positive cycles) this coincides with stable models; for non‑tight programs, ASPIC may admit additional supported models that are not answer sets. Integrating loop formula generation as in ASSAT of \cite{lin2004assat} is left as future work.
ASPIC has currently other limitations as well: support for explicit aggregating operators (\#min, \#max, \#count, \#sum) only in constraints, not yet in rules; experimental syntax for finite-domain variables (thus the syntax is subject to change).

We would argue that these limitations are only temporary: the technique of \cite{lin2004assat} can be used to break the circular reasoning, the syntax for finite-domain variables will eventually mature, when we and the users of ASPIC reach the best ways to benefit from combining ASP and Picat (including leveraging its support for constraint programming with finite-domain variables).

\section{Conclusion}

ASPIC, as a proof-of-concept transpiler, demonstrates that automatic conversion of ASP to Picat is possible and practical. The possibility of embedding Picat ``islands'' of code into ASP programs is barely explored, but we have used it already to add support for FD variables and to access Picat global constraints. ASPIC has shown promising results and has a modelling expressiveness advantage over pure clingo in certain scenarios, by tapping into Picat's CP and SAT solving strengths. It is also interesting that the SAT solver embedded into Picat can outperform even the multi-threaded clingo on some of our benchmarks (e.g. for the Towers of Hanoi encoding).

\section{Acknowledgments}
This research is partly supported by the project “Romanian Hub for Artificial Intelligence - HRIA”, Smart Growth, Digitization and Financial Instruments Program, 2021-2027, MySMIS no. 351416

\nocite{*}
\bibliographystyle{eptcs}
\bibliography{main} 

\end{document}